\documentclass[
    ,final            
  ]
  {aipproc}

\layoutstyle{8x11single}

\begin{document}

\title{Spectroscopic Binaries in Planetary Nebulae}


\author{Howard E. Bond}{
  address={Space Telescope Science Institute, 3700 San Martin Dr., Baltimore, MD
21218 USA}
}

%

\begin{abstract}

It is already known that about 10\% of central stars of PNe are very
short-period binaries (hours to days), which are detected through photometric
variations. These must have been formed through common-envelope interactions in
initially wide binaries, accompanied by ejection of the envelope and its
subsequent photoionization as a PN\null. Radial-velocity observations by
ourselves and others are now suggesting that an even larger fraction of
planetary nuclei may be spectroscopic binaries, making the total binary fraction
very large. However, we have not as yet been able to rule out the possibility
that the apparent velocity changes are actually due to stellar-wind variations.
Pending follow-up spectroscopic observations with large telescopes, it presently
appears  plausible that binary-star ejection is the major formation channel for
planetary nebulae.

\end{abstract}

\maketitle


\def\ace{\alpha_{\rm CE}}

\section{Are Most Planetary Nebulae Ejected from Binary Stars?}

In this paper, I report results of research by myself and my colleagues Orsola
De Marco, Andrew Fleming, Dianne Harmer, Robin Ciardullo, Melike Afsar, and Al
Grauer.

There are at least three arguments that suggest that many planetary nebulae
(PNe) are ejected from binary stars:

1. A large majority of PNe have highly non-spherical or bipolar shapes. The
simplest explanation of this fact would be that PN ejection occurs through
common-envelope (CE) interactions, in which the more massive star in an
initially wide binary evolves to red-giant dimensions and then engulfs its
main-sequence companion. The ensuing spiralling down of the orbit spins up the
envelope until it is ejected non-spherically, leaving a much closer binary
consisting of the hot core of the red giant (which photoionizes the ejected
envelope) and the companion. If there are PNe that were not actually ejected in
a CE interaction, the ejection process may still be strongly influenced by a
companion star, through tidal spin-up of the envelope and/or dynamo
magnetic-field generation.

2. There is an {\it observed\/} high incidence of {\it very close\/} binaries
among planetary-nebula nuclei (PNNi). Photometric monitoring programs have shown
that $\sim$10\% of PNNi are binaries with periods of a few hours to a few days
(e.g., several papers by Grauer \& Bond in the 1980's; Bond \& Livio 1990; Bond
2000). Such very close systems {\it must\/} have emerged from CE interactions.

The final orbital period following a CE interaction depends on the efficiency
with which orbital energy goes into ejecting material from the system, denoted
$\ace$. If $\ace$ is high, the final orbital periods will tend to be long, but
if $\ace$ is low, the periods will be shorter (and mergers of the cores will
occur more often).  Further discussion and theoretical simulations are given in
the following paper by De~Marco.

3. The photometric method for finding binary PNNi relies on the heating effect
of the hot nucleus on the facing hemisphere of the main-sequence companion,
which occurs only in very close binaries. Thus the 10\% of PNNi that can be
detected as binaries photometrically could be only the short-period tail of a
much larger overall binary fraction. Population-synthesis studies suggest that
this is in fact true for a wide range of $\ace$ values (e.g., Yungelson et al.\
1993).

Figure~1 shows the predicted period distribution for binary PNNi from Yungelson
et al.\ (with additional labels added by the present author) for a simulation
with an assumed high value of $\ace$.  There are three regimes labelled in the
diagram:  (a)~At long initial periods, the system does not enter into a CE, and
the period at the time a PN is visible is essentially unchanged (actually, it is
slightly longer due to the loss of material from the binary); when the
difference in magnitude between the components is small enough and the angular
separation large enough, such systems can be seen as resolved visual binaries. A
``snapshot'' search for visual binaries among PNNi was carried out with the {\it
Hubble Space Telescope\/} by Ciardullo et al.\ (1999), and resulted in the
discovery of 10 likely physical pairs out of 113 PNNi examined. These systems
are useful for determining distances to the PNe through main-sequence fitting,
but a significant effect of the binarity on the morphology of the nebula is
unlikely. (b)~At shorter periods, around 1000~days, Figure~1 predicts a lack of
binary PNNi, because such systems {\it did\/} enter into a CE interaction, and
ended up at shorter periods. (c)~The peak on the left side of Figure ~1 is due
to post-CE systems, which have spiralled down to short periods. For
$\ace\simeq1$, we see that the photometrically detectable binaries with periods
up to a few days could be only a fraction of the total, which extend up to
periods of more than 100~days.

\begin{figure}
  \includegraphics[width=4in]{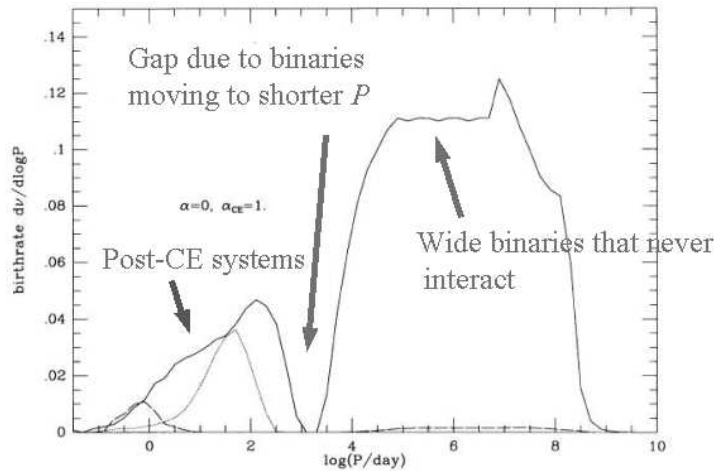}
  \caption{Theoretically predicted distribution of orbital periods among PN
central stars if $\ace=1$, from Yungelson et al.\ (1993). Added labels indicate,
from left to right,
systems that have emerged from common-envelope interactions, a gap where
systems are
absent because they have spiralled down to shorter periods, and wide systems
that never entered a CE.}
\end{figure}

Thus the $\sim$10\% of all PNNi known to be binaries in the photometrically
detectable regime could simply be the ``tip of the iceberg'' of a very large
overall binary fraction. This would mean that {\it PNe are fundamentally a
binary-star phenomenon}.

\section{Testing the ``Iceberg'' Hypothesis}

An observational test of this hypothesis requires radial-velocity (RV)
measurements, in order to find the wider binaries that lack photometrically
detectable heating effects.

My colleagues and I have been carrying out two spectroscopic surveys of PNNi in
order to detect periodic RV variations indicating binary membership:

1. In the northern hemisphere, De~Marco, Fleming, Harmer, and myself have used
the 3.5-m WIYN telescope and Hydra multi-object spectrograph at Kitt Peak to
monitor RV's.

2. And in the southern hemisphere, Afsar and myself are using the 1.5-m SMARTS
telescope and its Cassegrain spectrograph at Cerro Tololo.

\section{The WIYN Radial-Velocity Program}

With the WIYN spectrograph, we obtain a dispersion of 0.33~\AA/pix and a
spectroscopic resolution of $\sim$7000, leading to an RV precision of
$\sim$3--3.7~km/s. Our RV monitoring was carried out from 2002 to 2004, with
telescope scheduling optimized to find periods of days to months.

Results of the 2002-3 monitoring have been published by De Marco et al. (2004),
and can be summarized by saying that the fraction of PNNi with variable
velocities is extremely high. Specifically, 10 out of 11 PNNi with 4 to 16
measurements each have RV's that are variable, with statistical confidence
levels of 98 to 100\%.

\section{The SMARTS Radial-Velocity Program }

With the SMARTS 1.5-m spectrograph, observations were carried out in 2003-4, and
are being analyzed by Melike Afsar as part of her PhD thesis at Ege University,
Turkey. The dispersion is lower than at WIYN, 0.77~\AA/pix, giving a resolution
of $\sim$2000 and an RV precision of about 10~km/s. As with WIYN, the scheduling
was optimized to search for periods of days to months. 

Once again, we find strong evidence for variable RV's, in spite of the lower
precision. Out of 19 PNNi with at least 5 measurements, 7 have variable RV's
with greater than 99\% confidence. Reassuringly, 4 of the PNNi are in common
with those in the WIYN program; all 4 were found variable in the WIYN data, and
3 of them are also variable in the SMARTS data (with the fourth also being
variable at the 88\% confidence level). 

\section{Radial-Velocity Summary and Caveats}

At WIYN ($\sigma\sim3.5$~km/s) we found 10 out of 11 PNNi to have variable RV's,
and at SMARTS ($\sigma\sim10$~km/s) 7 out of 19 have variable RV's. 

Sorensen \& Pollacco (2004) have independently carried out an RV survey with a
precision of approximately 5~km/s, and found 13 out of 33 central stars to have
variable RV's (including NGC~6891, which we also find to be variable from both
WIYN and SMARTS data).

Do these results {\it prove\/} that the binary fraction is extremely high among
PNNi?  Unfortunately, it is too soon to assert this conclusion.

First, the RV measurements are difficult in some PNNi, because there are few
photospheric absorption lines, especially those free of nebular emission-line
contamination, and the velocity amplitudes we have found are generally low.
Moreover, it could be that some or many of the absorption lines contain
components due to a strong stellar wind, and that variations in these winds are
mimicking true RV variations. Although we did make an effort to select PNNi that
are relatively free of strong winds (e.g., we discriminated against stars with
strong P~Cygni features in their UV spectra), variable winds remain a possible
explanation for at least some of the apparent velocity variability.

Finding a {\it periodic\/} RV variation would greatly strengthen the binary
interpretation, but unfortunately we have not as yet been able to fit a
completely convincing period to any of our targets, with the possible exception
of the nucleus of IC~4593, which may have a 5.088~day period, as shown in
Figure~2.

\begin{figure}[b]
  \includegraphics[width=4in]{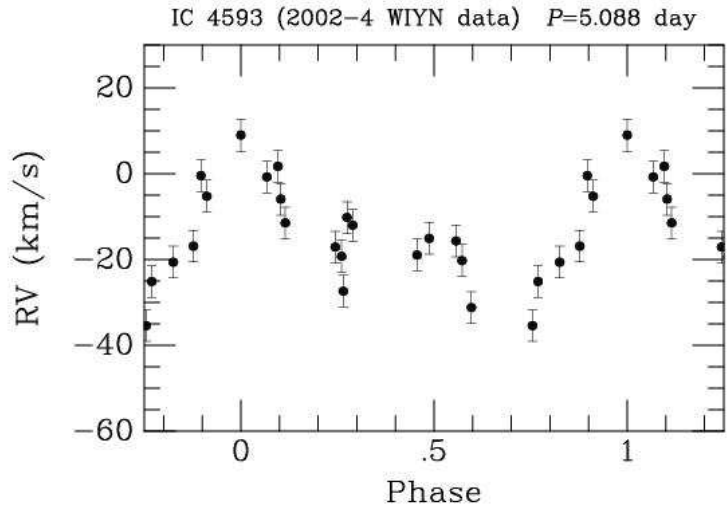}
  \caption{Radial velocities of the central star of IC 4593, measured with the
WIYN 3.5-m spectrograph and phased with a period of 5.088~days. If these
variations are due to true binary motion (rather than to stellar-wind
variations), the mass function is $f(m)=0.006$, implying $m_2>0.13M_\odot$ if
$m_1=0.6M_\odot$.}
\end{figure}

The problem is that our observations are indicating RV's that vary on a
timescale as short as one day, whereas our epochs were sparsely scattered over
an interval of 2-3 years. This is highly non-optimal for fitting the data with
periods that now appear to be as short as a few days.

\section{What Next?}

Based on the above results, we have concluded that the next step will be
intensive spectroscopic monitoring of a few PNNi for which we have already
detected RV variations, using a large telescope, high spectral resolution, and
high S/N\null. This will allow us to distinguish between wind variations (which
will produce variations in the line profiles) and true binary motion (in which
the entire profile will move back and forth periodically). De~Marco and I
recently used 5 nights on the Kitt Peak 4-m Mayall telescope and its echelle
spectrograph to carry out such observations, and the results will be reported in
the future.

\section{Implications of a High Binary Fraction among PN Nuclei }

There are several interesting astrophysical implications if the binary fraction
among PNNi is indeed very high.

One involves the existence of PNe in globular clusters, such as the famous K~648
in M15. The post-AGB remnants of low-mass stars are expected to evolve so slowly
across the HR diagram that, by the time they become hot enough to ionize an
ejected envelope, the envelope would already have dissipated into space. Thus,
the existence of 4 known PNe in globular clusters is difficult to understand.
However, binaries can merge or transfer matter before the PN stage is reached,
producing higher-mass remnants that do evolve fast enough. See Alves, Bond, \&
Livio (2000) for further discussion of this point.

Another point involves the existence of PNe at the bright end of the luminosity
function, which are used to determine extragalactic distances. These may be
descended from binaries, as discussed in a paper by Robin Ciardullo at this
meeting.

Most of the classes of compact binaries are probably descended from binary PNNi
via common envelopes. These include pre-cataclysmic binaries containing a red
dwarf and a white dwarf, exemplified by the well-known V471~Tauri in the Hyades
cluster (O'Brien, Bond, \& Sion 2001 and references therein), cataclysmic
variables, low-mass X-ray binaries, and SN~Ia progenitors.

Knowing the overall period distribution among PNNi would help constrain the
typical value of $\ace$, which is needed for population-synthesis calculations.

\begin{theacknowledgments}

In addition to the scientists thanked at the beginning, I acknowledge the
assistance of the service observers at the SMARTS 1.5m telescope (especially
Alberto Pasten and Sergio Gonzalez), and support from the STScI Director's
Discretionary Research Fund.

\end{theacknowledgments}

\bibliographystyle{aipprocl} 

\bibliography{sample}

\end{document}